\documentclass[aps,pre,preprint, showpacs, superscriptaddress]{revtex4}
\usepackage[dvips]{graphicx}
\usepackage{latexsym}
\usepackage{natbib}

\begin{document}
\title{Evolution equation for a model of surface relaxation in complex networks.}
\author{C.~E.~La~Rocca}
\affiliation{Instituto de Investigaciones F\'isicas de Mar del Plata (IFIMAR)-Departamento de F\'isica, Facultad de Ciencias Exactas y Naturales, Universidad Nacional de Mar del Plata-CONICET, Funes 3350, (7600) Mar del Plata, Argentina.}
\author{L. A. Braunstein}
\affiliation{Instituto de Investigaciones F\'isicas de Mar del Plata (IFIMAR)-Departamento de F\'isica, Facultad de Ciencias Exactas y Naturales, Universidad Nacional de Mar del Plata-CONICET, Funes 3350, (7600) Mar del Plata, Argentina.}
\affiliation{Center for Polymer Studies, Boston University, Boston, Massachusetts 02215, USA}
\author{P. A. Macri}
\affiliation{Instituto de Investigaciones F\'isicas de Mar del Plata (IFIMAR)-Departamento de F\'isica, Facultad de Ciencias Exactas y Naturales, Universidad Nacional de Mar del Plata-CONICET, Funes 3350, (7600) Mar del Plata, Argentina.}
\begin{abstract}
In this paper we derive analytically the evolution equation of the interface
for a model of surface growth with relaxation to the minimum (SRM) in complex
networks.  We were inspired by the disagreement between the scaling results of
the steady state of the fluctuations between the discrete SRM model and the
Edward-Wilkinson process found in scale-free networks with degree distribution
$ P(k) \sim k^{-\lambda}$ for $\lambda <3$ [Pastore y Piontti {\it et al.}, Phys. Rev. E {\bf 76}, 046117 (2007)]. Even though for Euclidean lattices the
evolution equation is linear, we find that in complex heterogeneous networks
non-linear terms appear due to the heterogeneity and the lack of symmetry of
the network; they produce a logarithmic divergency of the saturation roughness
with the system size as found by Pastore y Piontti {\it et al.} for $\lambda <3$.
\end{abstract}


\maketitle

During the last few years the study of complex networks has moved its focus
from the study of their topology to the dynamic processes occurring on the
underlying network. This is because many physical and dynamic processes use
complex networks as substrates. Recently, many studies of dynamic processes on
networks, such as epidemic spreading \cite{Pastorras_PRL_2001}, traffic flow
\cite{Lopez_transport,zhenhua}, cascading failure \cite{Motter_prl}, and
synchronization \cite{Jost-prl,Korniss07}, have demonstrated the importance of
the topology of the substrate network in the dynamic process. There exists
much evidence that many real networks possess a scale-free (SF) degree
distribution characterized by a power law tail given by $ P(k) \sim
k^{-\lambda}$, where $k_{max} \ge k \ge k_{min}$ is the degree of a node,
$k_{max}$ is the maximum degree, $k_{min}$ is the minimum degree, and
$\lambda$ measures the broadness of the
distribution~\cite{Barabasi_sf}. Almost all the studies on networks regarded
the links or nodes as identical. However, in real networks the links or nodes
are not identical but have some ``weight.'' As examples the links between
computers in the internet network have different capacities or bandwidths,
resistor networks can have different values of resistance \cite{zhenhua}, and
the airline network links connecting pairs of cities in direct flights have
different numbers of passengers. Many theoretical studies have been carried
out on weighted networks \cite{weigthed-papers,zhenhua}.  Recently, several
studies on real networks with weights on the links, such as the world-wide
airport networks and the {\it Escherichia coli} metabolic
networks~\cite{Barrat_pnas}, have shown that the weights are correlated with
the network topology and this dramatically changes the transport through them
\cite{Zhenhua_corr,Korniss07}. For instance, in synchronization problems,
which are very important in brain networks \cite{JWScanell}, networks of
coupled populations in the synchronization of epidemic outbreaks
\cite{eubank_2004}, and the dynamics and fluctuations of task completion
landscapes in causally constrained queuing networks \cite{Kozma05}, the
weights could have dramatic consequences for the synchronization
\cite{Korniss07}.  Synchronization problems deal with optimization of the
fluctuations of some scalar field $h$. The system will be optimally
synchronized when the fluctuations are minimized.  The general treatment to
analyze the fluctuations of these processes is to map them into a problem of
non-equilibrium surface growth via an Edwards-Wilkinson (EW) process on the
corresponding network \cite{EW}.  Given a scalar field $h$ on the nodes, that
represents the interface height at each node, the fluctuations are
characterized by the average roughness $W(t)$ of the interface at time $t$,
given by $W \equiv W(t) = \left\{1/N \sum_{i=1}^N (h_i-\langle h
\rangle)^2\right\}^{1/2},$ where $h_i \equiv h_i(t)$ is the height of node $i$
at time $t$, $ \langle h \rangle$ is the mean value on the network, $N$ is the
system size, and $\{ . \}$ denotes an average over configurations. The EW process
on networks is given by
\begin{equation}\label{Eq.ew}
  \frac{\partial h_i}{\partial t}= \sum_{j=1}^N C_{i j}  (h_j -h_i ) + \eta_i\ ,
\end{equation}
where $C_{i j}= A_{i j}\ w_{i j}$ is a symmetric coupling strength, $\{
A_{ij}\}$ is the adjacency matrix ($A_{ij}=1$ if $i$ and $j$ are connected and
zero otherwise), $w_{i j}$ is the weight on the edge connecting $i$ and $j$,
and $ \eta_i (t)$ is a Gaussian uncorrelated noise with zero mean and
covariance $\{\eta_i \eta_j \} = 2 D \delta_{i j} \delta(t-t^{'})$. Here $D$
is the diffusion coefficient and is taken in general as a constant.  For
non-weighted networks $w_{i j}=\nu=\mbox{const}$ and thus Eq.~(\ref{Eq.ew})
reduces to the unweighted EW equation on a graph given by $\partial h_i
/\partial t= \nu \sum_{j=1}^N A_{i j} (h_j -h_i ) + \eta_i$.  Inspired by the
results found for real networks where the weights are correlated with the
topology, Korniss \cite{Korniss07} studied synchronization for EW
processes [see Eq.~(\ref{Eq.ew})] on SF networks where $w_{i j} =(k_i k_j)
^\beta$ and $k_i$ and $k_j$ are the degrees of the nodes connected by a link.
Using a mean-field approximation, he found that, subject to a fixed total edge
cost, synchronization is optimal when $\beta=-1$, and at that point the
performance is equivalent to that of the complete graph with the same edge
cost. Pastore y Piontti {\it et.  al} \cite{anita} used a discrete growth
model with surface relaxation to the minimum (SRM) in SF networks, which mimics
the fluctuation in the task-completion landscapes in certain distributed
parallel schemes on computer networks, because it balances the load. They
found that in SF networks with $\lambda <3$ the saturation regime of $W \equiv
W_s$ has a logarithmic divergence with $N$ that cannot be explained with the
unweighted EW equation in graphs, even though in Euclidean lattices the SRM
model belongs to the same universality class as the EW equation
\cite{family}.

In order to understand this discrepancy, in this paper
we derive analytically the evolution
equation for the SRM in random unweighted networks \cite{anita}
and find that the dynamics introduces ``weights'' on the links. With our
evolution equation, which contains non-linear terms in the height differences, we
recover the logarithmic divergency of $W_s$ with $N$ found in \cite{anita} for SF networks with $\lambda <3$.
Let us first briefly recall the SRM
discrete model \cite{family}, studied for SF networks by Pastore y Piontti {\it et. al} \cite{anita}.
In this model, at each time step a node $i$ is chosen with
probability $1/N$. If we denote by $v_i$ the nearest-neighbor nodes of $i$
and $j \in v_i$, then (1) if $h_i \leq h_j$ $\forall j \in v_i$ $\Rightarrow h_i= h_i+1$, else
(2) if $h_j < h_n$ $\forall n \not= j \in v_i$  $\Rightarrow
h_j = h_j+1$. Next we derive the analytical evolution equation for the local
height of the SRM model in random graphs. The
procedure chosen here is based on a coarse-grained (CG) version of the discrete
Langevin equations obtained from a Kramers-Moyal expansion of the master
equation \cite{VK,Vveden,lidia}. The discrete Langevin equation for the evolution of the
height in any growth model is given by \cite{Vveden,lidia}
\begin{eqnarray}\label{eqh}
\frac{\partial h_i}{\partial t}= \frac{1}{\tau}G_i + \eta_i,
\end{eqnarray}
where $G_i$ represents the deterministic growth rules that cause evolution of
the node $i$, $\tau=N \delta t$ is the mean time to grow a layer of the
interface, and $\eta_i$ is a Gaussian noise with zero mean and covariance given
by \cite{Vveden,lidia}
\begin{equation}\label{ruido}
\{\eta_i(t)\eta_j(t')\}=\frac{1}{\tau}G_i\delta_{ij}\delta(t-t')\ .
\end{equation}
We can write $G_i$ more explicitly as
\begin{equation}\label{eqreglas}
G_i = \omega_i + \sum_{j=1}^{N} A_{i j}\ \omega_j\ ,
\end{equation}
where $\omega_i$ is the growth contribution by deposition on  node $i$ and
$\omega_j$ is the growth contribution to node $i$  by relaxation  from any of its $j$
neighbors with
\[ \omega_{i}= \prod_{j \in v_i} \Theta(h_{j}-h_{i}),\]
\[ \omega_j= \left[ 1- \Theta(h_{i}-h_{j}) \right] \prod_{n \in v_j}
            \left[ 1- \Theta(h_{i}-h_{n}) \right]\ .\]
Here, $\Theta$ is the Heaviside function given by $\Theta(x)=1$ if
$x \geq 0$ and zero otherwise, with $x=h_t-h_s \equiv \Delta h$. Without lost
of generality, we take $\tau=1$ and
assume that the initial configuration of $\{h_i\}$ is random.

In the CG version $\Delta h \rightarrow 0$; thus after expanding an analytical
representation of $\Theta(x)$ in Taylor series around $x=0$ to second order in
$x$, we obtain
\begin{eqnarray}\label{nonliner}
G_{i}&&=\ c_0^{k_i}\ +\ C_i\ +\ c_1c_0^{k_i-1} k_i\left[\sum_{j=1}^{N}\frac{A_{ij}h_j}{k_i}-h_i\right]\nonumber\\
  &&+ \frac{c_1}{(1-c_0)}C_i\left[\sum_{j=1}^{N}\frac{C_{ij}h_j}{C_i}-h_i\right]
  +\
  \frac{c_1}{(1-c_0)}T_i\left[\sum_{j=1}^{N}\sum_{n=1,n\not=i}^{N}\frac{T_{ijn}h_n}{T_i}-h_i\right]\nonumber\\
  &&-\ c_2\ \sum_{j=1}^{N}A_{ij}\Omega(k_j-1)(h_j-h_i)^2\nonumber\\
  &&-\ \left[c_2+\frac{c_1^2}{2(1-c_0)}\right]\
  \sum_{j=1}^{N}A_{ij}\Omega(k_j-1)\left[\sum_{n=1,n\not=i}^{N}A_{jn}(h_n-h_i)^2\right]\nonumber\\
  &&+\ c_0^{k_i-1}\left[c_2-\frac{c_1^2}{2c_0}\right]\sum_{j=1}^NA_{ij}(h_j-h_i)^2\ +\ \frac{c_0^{k_i-2}c_1^2}{2}\left[\sum_{j=1}^NA_{ij}(h_j-h_i)\right]^2\nonumber\\
  &&+\
  \frac{c_1^2}{(1-c_0)}\sum_{j=1}^{N}A_{ij}\Omega(k_j-1)(h_j-h_i)\left[\sum_{n=1,n\not=i}^NA_{jn}(h_n-h_i)\right]\nonumber\\
  &&+\ \frac{c_1^2}{2(1-c_0)}\sum_{j=1}^{N}A_{ij}\Omega(k_j-1)\left[\sum_{n=1,n\not=i}^NA_{jn}(h_n-h_i)\right]^2\ ,
\end{eqnarray}
where $c_0$, $c_1$, and $c_2$ are the first three coefficients of
the expansion of $\Theta(x)$, $\Omega(k_j)=(1-c_0)^{k_j}$ is the weight on the link $ij$
introduced by the dynamic process, and
\begin{eqnarray}\label{peso1}
C_i=&\sum_{j=1}^{N}C_{ij}\ ;\nonumber\\
T_i=&\sum_{j=1}^{N}\sum_{n=1,n\not=i}^{N}T_{ijn}\ ,
\end{eqnarray}
with $C_{ij}=A_{ij}\Omega(k_j)$
and $ T_{ijn}=A_{ij}A_{jn}\Omega(k_j)$.

In our equation the non-linear terms in the difference of heights arise as a
consequence of the lack of a geometrical direction and the heterogeneity of
the underlying network. This result is very different from the one found in
Euclidean lattices, where for the
SRM model the non-linear terms disappear due to the symmetry of
the process and the homogeneity of the lattice.

For the noise correlation [see Eq.~(\ref{ruido})], up to zero
order in $\Delta h$ \cite{Vveden,lidia} we obtain $\{\eta_i(t)
\eta_j (t^{'})\}= 2 D(k_i)\delta_{ij} \delta(t-t')$ with
\begin{eqnarray}\label{coefdifus}
D(k_i)=\frac{1}{2}(c_0^{k_i} + C_i).
\end{eqnarray}

Notice that all the coefficients of the equation depend on the connectivity of
node $i$, {\it i.e.}, on the network topology of the underlying network. This
dependence on the topology can be thought of as a weight on the links of the
unweighted underlying network that appears only due to the dynamics on the
heterogeneous network.

\begin{figure}
\includegraphics[width=0.4\textwidth]{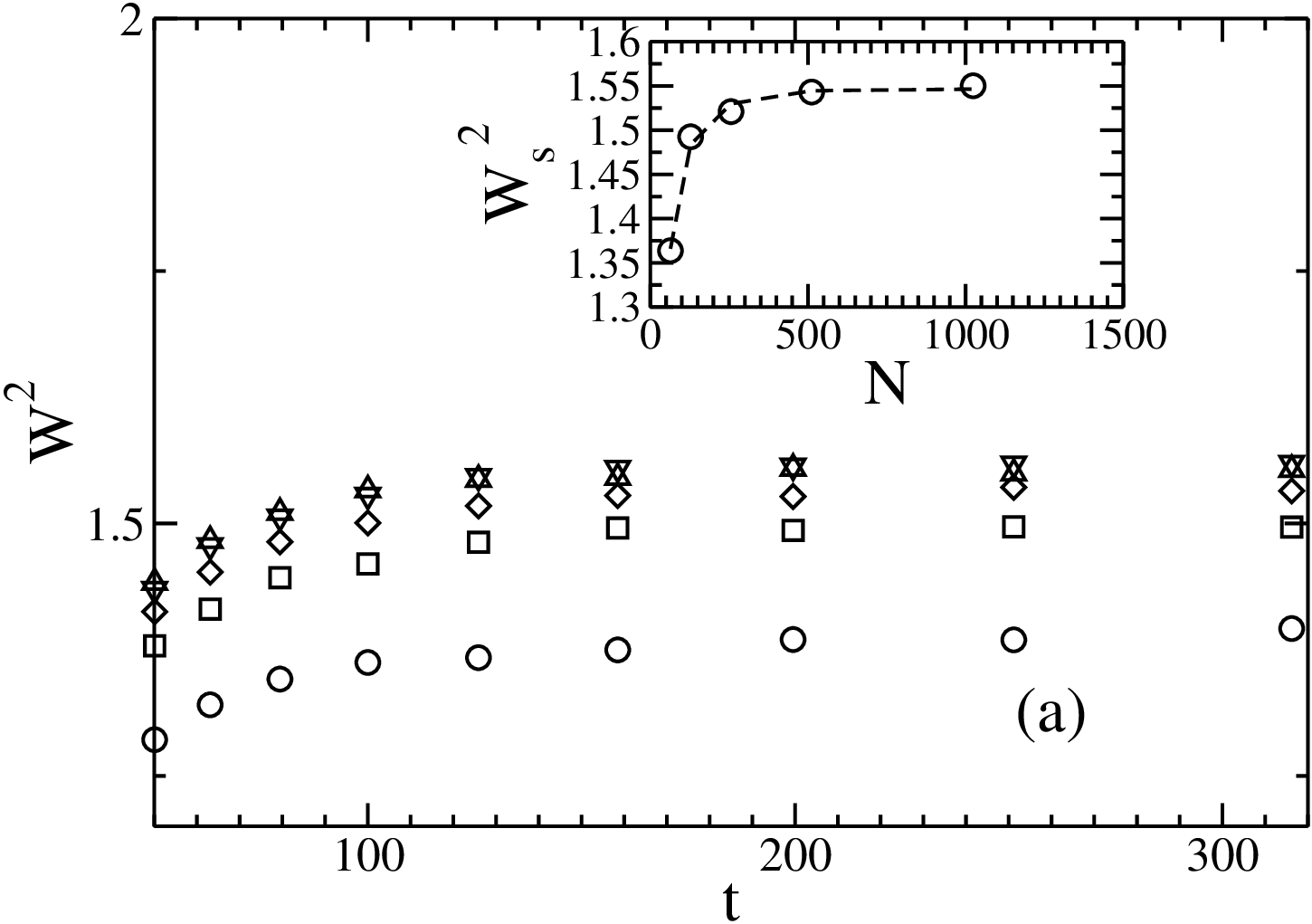}\\
\vspace{2cm}
\includegraphics[width=0.4\textwidth]{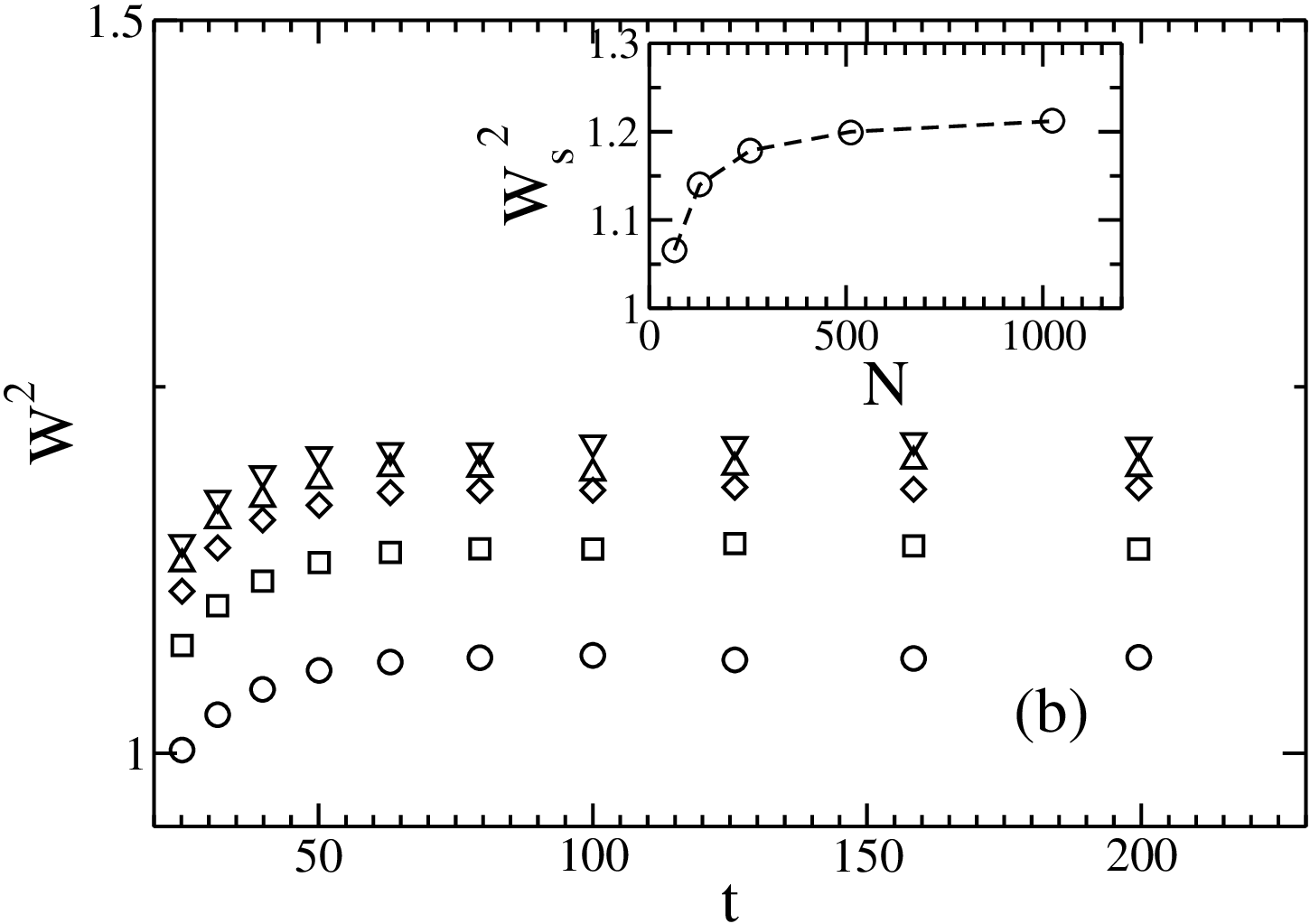}
\vspace{2cm}
\caption{$W^2$ as a function of $t$ from the integration of the evolution equation using the linear terms
 for $N=256$ ($\circ$), $384$ ($\Box$), $512$
($\diamond$), $768$ ($\bigtriangleup$), and $1024$
($\bigtriangledown$). $\lambda=(a)\ 3.5$ and $(b)\ 2.5$. In the inset figure we plot $W_s^2$ vs $N$ in
symbols. The dashed lines represent the fitting with
Eq.~(\ref{tamfinito}), obtained by
considering the finite-size effects introduced by the MR
construction. For all the integrations
we used $U= 0.5$ and typically $10\ 000$ realizations of
networks.\label{fig.1}}
\end{figure}

Interestingly, the linear terms are different from the EW process
as shown below. Keeping only the linear terms in
Eq.~(\ref{nonliner}), we numerically integrate our evolution
equation in a SF network using the Euler method with the
representation of the Heaviside function given by $\Theta(x)=\{1+
\tanh[U(x+z)]\}/2$, where $U$ is the width and $z= 1/2$
\cite{lidia}. With this representation $c_0=[1+\tanh(U/2)]/2$,
$c_1=[1-\tanh^2(U/2)]\ U/2$, and $c_2=[-\tanh(U/2)+\tanh^3(U/2)]\ U^2/2$. We
build the network using the Molloy-Reed (MR)
algorithm \cite{Molloy}. In Fig.~\ref{fig.1}, we plot $W^2$ as
a function of $t$, obtained from the integration of Eq.~(\ref{eqh})
using only the linear terms of Eq.~(\ref{nonliner}) with $D(k_i)$ given by
Eq.~(\ref{coefdifus}) for $\lambda=3.5$ and $2.5$ and
different values of $N$ with $k_{min}=2$ in order to ensure that the network
is fully connected. For the time step integration we chose $\Delta t \ll
1/k_{max}$ according to Ref.~\cite{pasointeg}. In contrast to the results obtained for
the EW process \cite{anita}, $W_s$ increases
with $N$ until it reaches a constant value. As shown below,
this dependence of $W_s$ on $N$ is due to finite-size effects
due to the MR construction.

Now we apply a mean-field approximation to the linear terms of
Eq.~(\ref{nonliner}). In this
approximation we consider $1 \ll k_{min} \ll k_{max}$ and
disregard the fluctuations. Then $ \sum_{j=1}^{N} A_{ij}h_j /k_i
\approx \langle h \rangle$, $ \sum_{j=1}^{N}C_{ij}h_j/C_i \approx
\langle h \rangle$, and
$\sum_{j=1}^{N}\sum_{n=1,n\not=i}^{N}T_{ijn}h_n/T_i \approx
\langle h \rangle $.  Multiplying and dividing Eq.~(\ref{peso1}) by $k_i$, we can
approximate $C_i$ by $C_i(k_i) \approx
k_i\int_{k_{min}}^{k_{max}} P(k|k_i)\ \Omega(k)\ d k$ \cite{Korniss07}, where $P(k|k_i)$ is the
probability that a node with degree $k_i$ is connected to another with degree
$k$. For uncorrelated networks, $P(k|k_i) = k P(k) /\langle k \rangle$
\cite{Barabasi_sf} does not depend on $k_i$; then $C_i(k_i)
\approx I_1\ k_i/\langle k\rangle$ with $I_1=\int_{k_{min}}^{k_{max}}
P(k)\ k\ \Omega(k)\ dk$ .  Making the same assumption for $T_i$,
we obtain $T_i(k_i) \approx
I_2\ k_i/\langle k\rangle$ with $I_2=\int_{k_{min}}^{k_{max}} P(k)\ k\ (k-1)\ \Omega(k)\ dk$.
Then the linearized evolution equation for the heights can be written as
\begin{eqnarray}\label{eqaprox}
\frac{\partial{h_i}}{\partial t}&=& F_i(k_i)+ \nu_i(k_i)\ (\langle h
\rangle-h_i) + \eta_i\ ,
\end{eqnarray}
where $F_i(k_i)=c_0^{k_i}+k_i\ I_1/\langle k \rangle$ represents a
local driving force, $\nu_i(k_i)=(c_1c_0^{k_i-1}+b)k_i$
is a local superficial tension-like coefficient with
$b=c_1(I_1 + I_2)/\langle k \rangle$, and $\eta_i$ is a
Gaussian noise with covariance $D(k_i)=F_i(k_i)/2$.
This approximation shows the full topology of the network through $P(k)$. 

Taking the average over the network in
Eq.~(\ref{eqaprox}), $\partial \langle h \rangle/\partial
t=1/N\sum_{i=1}^N F_i=F$;
then $\langle h \rangle=Ft$ is linear with $t$.
The solution of Eq.~(\ref{eqaprox}) \cite{VK} is given by
\begin{eqnarray}\label{solucion}
  h_i(t)&=& \int_{0}^{t}e^{-\nu_i(t-s)}\ (F_i+\nu_i\langle h(s)
  \rangle+\eta_i(s))\ ds\nonumber \\
 &=& \left(\frac{F_i-F}{\nu_i}\right) -\left(\frac{F_i-F}{\nu_i}\right)e^{-\nu_it} +
  \langle h \rangle + \int_{0}^{t}e^{-\nu_i(t-s)}\eta_i(s)\ ds\ .
\end{eqnarray}

\begin{figure}
\includegraphics[width=0.4\textwidth]{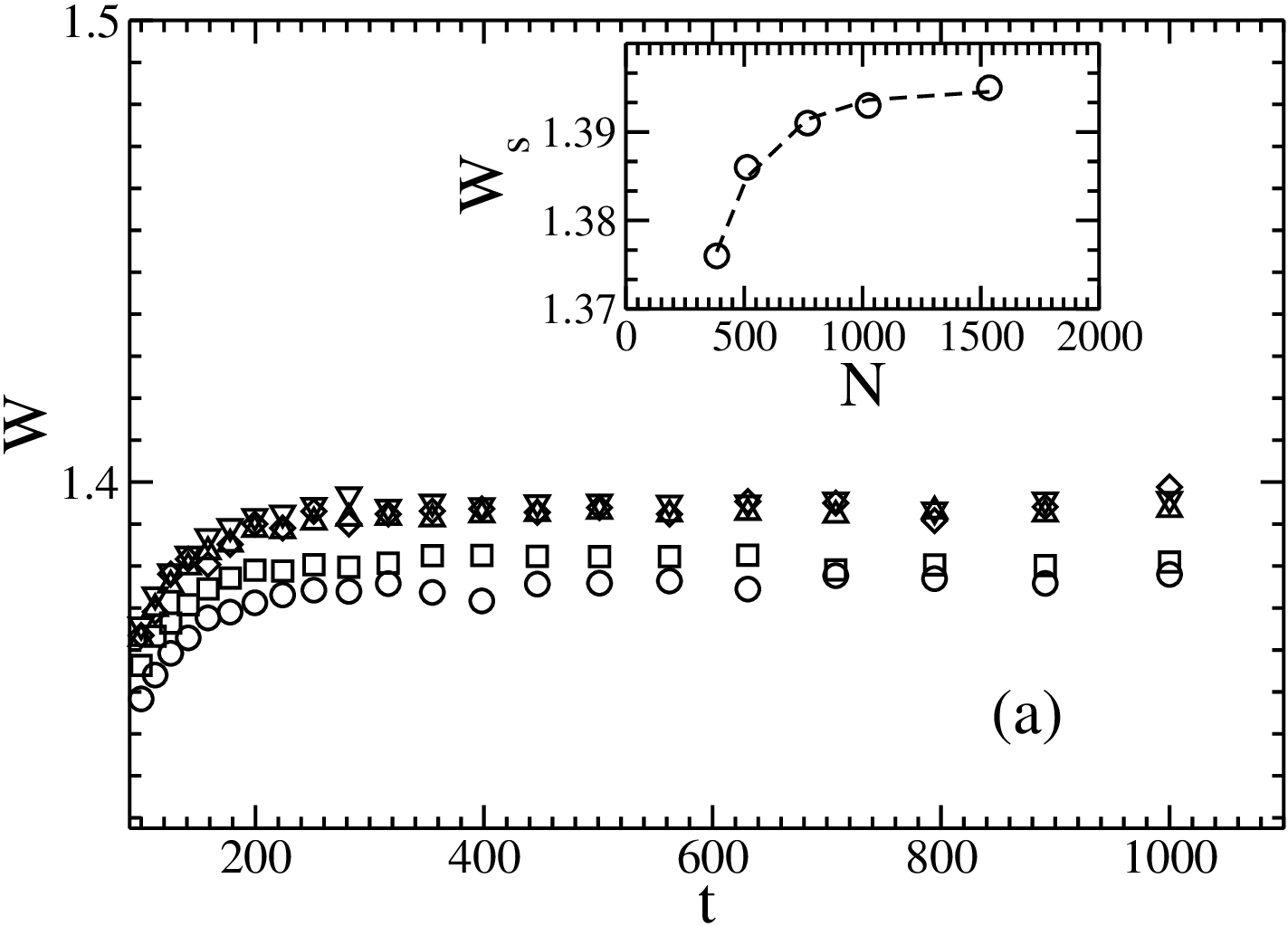}\\
\vspace{2cm}
\includegraphics[width=0.4\textwidth]{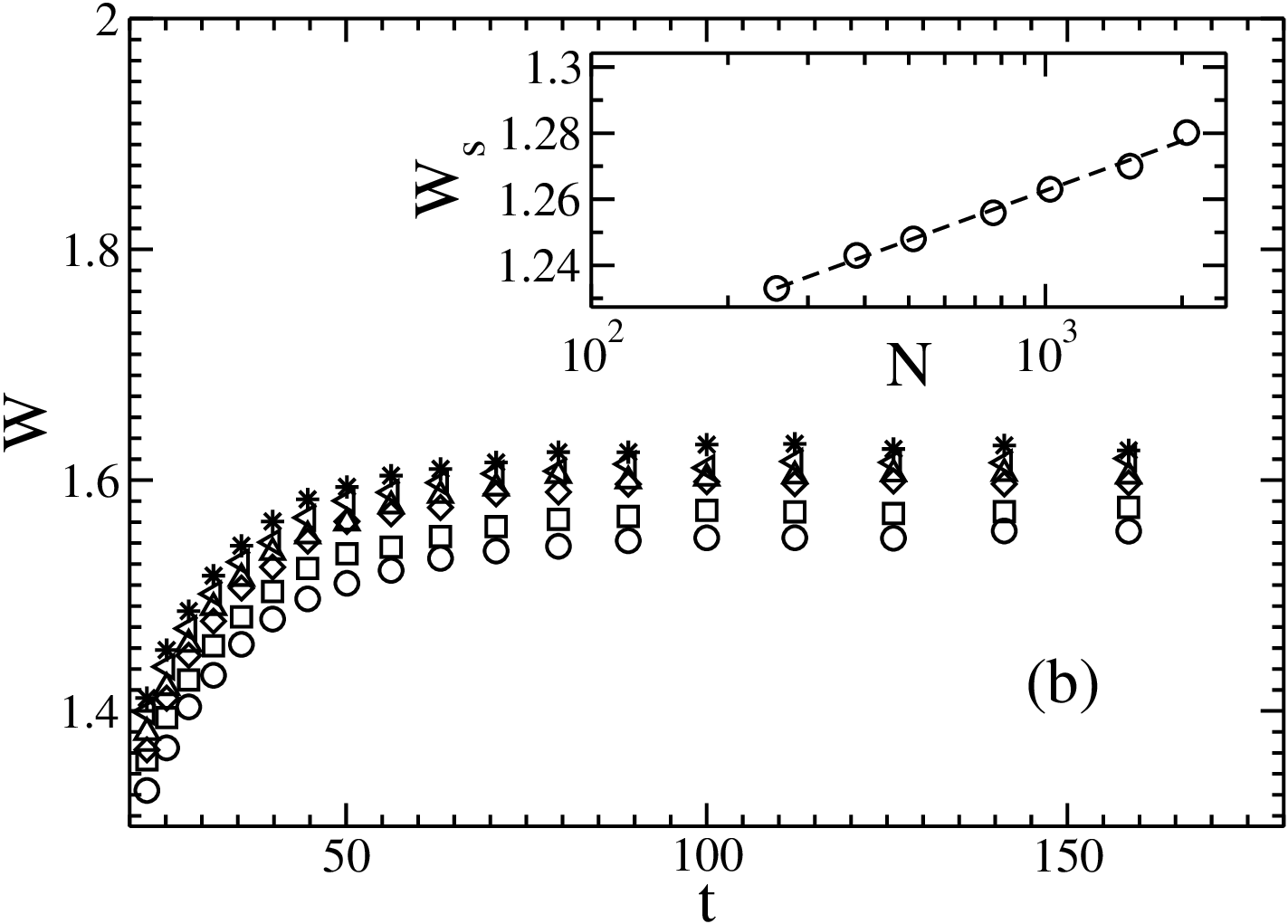}\vspace{2cm}
\caption{$W$ as a function of $t$ from the integration of the evolution equation
using the linear terms and the first non-linear term for $N=384$ ($\circ$),
$512$ ($\Box$), $768$ ($\diamond$), $1024$ ($\bigtriangleup$), and $1536$
($\bigtriangledown$). $\lambda=(a)\ 3.5$ and $(b)\ 2.5$. In the inset
figure we plot $W_s$ vs $N$ in symbols. The dashed lines represent the fitting
with Eq.~(\ref{tamfinito}) in $(a)$ and $W_s \sim \ln N$ in $(b)$.\label{fig.2}}
\end{figure}

Using Eq.~(\ref{solucion}), the two-point correlation function for
$t>max\left\{1/\nu_i \right\} \sim 1/k_{min}$, is
\begin{eqnarray*}
\left\{(h_i(t_1)-\langle h \rangle)(h_j(t_2)-\langle h
  \rangle)\right\}&&=\ \ \left(\frac{F_i-F}{\nu_i}\right)\left(\frac{F_j-F}{\nu_j}\right)\\ \nonumber
  +&& \int_{0}^{t_2}\int_{0}^{t_1}e^{-\nu_i(t_1-s_1)}e^{-\nu_j(t_2-s_2)}\left\{\eta_i(s_1)\eta_j(s_2)\right\}\
  ds_1ds_2\ .\nonumber
\end{eqnarray*}
Then $W_s$ can be written as
\begin{equation}\label{rugsat}
W_s^2=\frac{1}{N}\sum_{i=1}^N\left(\frac{F_i-F}{\nu_i}\right)^2
  + \frac{1}{N}\sum_{i=1}^N\frac{2D(k_i)}{2\nu_i}\ .
\end{equation}
For SF networks it can be shown that $I_1, I_2 \sim
\mbox{const.}+k_{max} \exp(-k_{max}\mbox{const.})$, where
$k_{max}\sim N^{1/(\lambda -1)}$ for MR networks; thus finite-size
effects due to the cutoff on these quantities can be disregarded.
Replacing in the last equation $D(k_i)$ by
$F_i(k_i)/2$, we obtain

\begin{equation}\label{rugescal}
W_s^2\sim \left[1-2\langle k\rangle\  \langle \frac{1}{k}\rangle + \langle
  k\rangle^2\  \langle \frac{1}{k^2}\rangle \right]+\mbox{const.}\ .
\end{equation}
Notice that, if $F_i=0$, $D=\mbox{const}$, and $\nu_i\propto
k_i$, we recover the EW equation found in
\cite{Korniss07}. Using the corrections due to finite-size effects introduced by
$k_{max}$ \cite{anita} in Eq.~(\ref{rugescal}),
\begin{equation}\label{tamfinito}
W_s^2\sim W_s^2(\infty)\left[1+q_1\frac{1}{N^{\frac{\lambda -2}{\lambda
    -1}}}+q_2\frac{1}{N}\right]\ ,
\end{equation}
where $W_s^2(\infty)=W_s^2(N \to \infty)$ and $q_1$ and $q_2$ are
constants. In the inset of Figs.~\ref{fig.1} $(a)$ and $(b)$ we plot $W_s^2$
as function on $N$ and the fitting obtained from Eq.~(\ref{tamfinito}). The
agreement with the scaling form, Eq.~(\ref{tamfinito}), is excellent. Thus,
the linear approximation can only explain the finite-size effects due to the
MR construction but fails to predict the logarithmic divergency of $W_s$ with
$N$ for $\lambda <3$ found in Ref.~\cite{anita}. Next we show that the
non-linear terms are responsible for this behavior.  We integrate our
evolution equation for SF networks with the linear terms and only the first
non-linear term [see Eq.~(\ref{nonliner})] due to the numerical instability
produced when we try to incorporate all of them. Even with only one non-linear
term, we recover the logarithmic divergency of $W_s$ with $N$ for $\lambda
<3$. The results of the integration are shown in Fig.~\ref{fig.2}, where we
plot $W$ as a function of $t$ for (a) $\lambda=3.5$ and (b) $\lambda=2.5$ and
different values of $N$. In the inset figures we plot $W_s$ as a function of
$N$. We can see that, for $\lambda=3.5$, $W_s$ increases but asymptotically
goes to a constant and all the $N$ dependence is due to finite-size
effects. However, for $\lambda=2.5$ we found a logarithmic divergency of
$W_s$ with $N$ \cite{anita}, as shown in the inset of Fig.~\ref{fig.2} $(b)$,
where we plot $W_s$ as a function of $N$ on a log-linear scale. The fit of $W_s$
with a logarithmic function for $\lambda=2.5$ shows the agreement between our
results and those obtained for the SRM model in SF networks for $\lambda <3$.
Discrepancies between behaviors in regular Euclidean lattices and Euclidean
lattices after addition of random links were found before in \cite{refereesnote}.

In summary, we derived analytically the evolution equation for the SRM model
and found, surprisingly, that even when the underlying network is unweighted
the dynamics introduces weights on the links that depend on the topology. We
also found that the linear terms can explain only finite-size effects due to
the MR construction. The linear mean-field approximation shows clearly the
effects of the topology on the dynamics and the corrections due to finite-size
effects.  When non-linear terms on SF networks are considered, new numerical
integration algorithms are needed in order to avoid numerical
instabilities. This is still an open problem to be solved in the future.  With
all the linear terms and one non-linear term, we recovered the logarithmic
divergency of $W_s$ with $N$ of the SRM model for $\lambda<3$. Our analytic
procedure can be also applied to any other growth model.

We thank A. L. Pastore y Piontti for useful discussions and comments. This
work has been supported by UNMdP and FONCyT (Pict 2005/32353).

\end{document}